%% file: SEBNAP.tex
\documentclass[superscriptaddress,showpacs,amsmath,amssymb,twocolumn,nofootinbib]{revtex4-2}

\usepackage{amsmath,amsthm,amssymb,bm,braket}
\usepackage[dvips]{graphicx}
\usepackage{color}
\usepackage{wrapfig}
\usepackage{fancybox}

\newcommand{\tomo}[1]{\textcolor{black}{#1}}

\begin{document}
\title{One-proton emission of $^{102}$Sb and its sensitivity to proton-neutron interaction}

\author{Tomohiro Oishi}
\email[E-mail: ]{tomohiro.oishi@ribf.riken.jp}
\affiliation{RIKEN Nishina Center for Accelerator-Based Science, Wako 351-0198, Japan}

\author{Masaaki Kimura}
\email[E-mail: ]{masaaki.kimura@ribf.riken.jp}
\affiliation{RIKEN Nishina Center for Accelerator-Based Science, Wako 351-0198, Japan}

\author{Lorenzo Fortunato}
\email[E-mail: ]{lorenzo.fortunato@pd.infn.it}
\affiliation{Department of Physics and Astronomy ``Galileo Galilei'', University of Padova, and I.N.F.N. Sezione di Padova, via F. Marzolo 8, IT-35131, Padova, Italy.}

\renewcommand{\figurename}{FIG.}
\renewcommand{\tablename}{TABLE}

\newcommand{\bi}[1]{\ensuremath{\boldsymbol{#1}}}
\newcommand{\unit}[1]{\ensuremath{\mathrm{#1}}}
\newcommand{\oprt}[1]{\ensuremath{\hat{\mathcal{#1}}}}
\newcommand{\abs}[1]{\ensuremath{\left| #1 \right|}}

\def \beq{\begin{equation}}
\def \eeq{\end{equation}}
\def \beqa{\begin{eqnarray}}
\def \eeqa{\end{eqnarray}}
\def \Schr{Schr\"odinger }
\def \pn{{\it pn}}

\def \bir{\bi{r}}
\def \ubir{\bar{\bi{r}}}
\def \bip{\bi{p}}
\def \ubip{\bar{\bi{r}}}

\begin{abstract}
One-proton emission from the $^{102}$Sb nucleus is discussed, assuming an inert $^{100}$Sn core and the valence proton and neutron.
There are experimentally measured bound states in the $^{100}$Sn-neutron system, whereas no particle-bound $^{100}$Sn-proton state has been observed.
With time-dependent three-body calculations, the $1^+$ ground state of $^{102}$Sb is suggested as a possible proton emitter.
This conclusion is reached by assuming a weakening effect on the proton-neutron ($pn$) interaction with respect to a bare deuteron.
An analogous phenomenon is necessary to reproduce the empirical binding energies of $^{42}$Sc and $^{18}$F.
Continuous shift from the unbound to bound regions by changing the $pn$-interaction strength is demonstrated.
The lower limit of lifetime is evaluated as $\tau \gtrsim 4.4 \times 10^{-18}$ seconds in the no-$pn$-interaction limit.
However, the actual lifetime is expected as longer with a finite $pn$ interaction.
Observation of a resonant state in $^{102}$Sb and its decay would provide a benchmark of the $pn$-pairing correlation.
\end{abstract}

\maketitle

\section{Introduction} \label{Sec:intro}
Proton-rich nuclei around $A=100$ have attracted a special interest in recent years \cite{2016Tani,2014Kalanda,2014Doornenbal,2015Maharana,2019Xu}.
A variety of radioactive nuclides, including alpha, proton, and light-particle emitters, can exist in this region, due to the proximity of the dripline.
In addition, the $rp$ process, which contains rapid captures of protons followed by the $\beta^+$ decay, is expected along the $N\cong Z$ line in order to explain the origin of proton-rich isotopes.
For such interests, the $^{102}$Sb nucleus is one enchilada.
Whether this system is particle-bound or not is an open question.

For the $N \cong 50$ and $Z \cong 50$ nuclei,
the Coulomb repulsion of protons is relatively strong, and many systems are expected as particle-unbound.
On the other hand, as the counter effect to Coulomb repulsion,
the proton-neutron ($pn$) interaction can stabilize these nuclei.
When the valence protons and neutrons occupy the same orbits, the $pn$-attractive effect is enhanced.
The $^{102}$Sb nucleus is one typical case.
Considering $^{100}$Sn as the rigid core, the core-neutron subsystem $^{101}$Sn is bound with
$S_n = 11.2(4)$ MeV \cite{NNDC_Chart} (one-neutron separation).
In contrast, the core-proton subsystem $^{101}$Sb is expected as particle-unbound.
By assuming a strong $pn$ attraction, the particle-binding nature of $^{100}$Sn-$n$-$p$ is expected. 
Therefore, $^{102}$Sb provides a good benchmark to infer the $pn$-pairing strength as well as the deuteron correlation \cite{2016Tani,2012Tani,2016Masui,2024Uzawa}.
In Ref. \cite{2016Tani}, the effect of isoscalar spin-triplet $pn$ pairing in $^{102}$Sb is deeply investigated.
The charge-exchange Gamow-Teller transitions are shown to be significantly sensitive to this $pn$ pairing.
On the other hand, the alternative scenario of unbound $^{102}$Sb has been less considered.
Especially the one-proton emission is expected if the $pn$ interaction is not strong.

Proton-rich nuclei around $A = 100$ have been experimentally produced by the
(i) multi-nucleon transfer \cite{2019Xu,2015Devaraja,2020Kalandarov},
(ii) fragmentation of neutron-deficient nuclei \cite{1994Hencheck,1995Rykacz,2016Celikovic}, and
(iii) fusion-evaporation reaction \cite{1996Chartier,2000Commara,2009Elomaa,2018Auranen}.
For example, in Ref. \cite{2015Devaraja}, a multi-nucleon-transfer reaction $^{48}$Ca$+^{248}$Cm was utilized to produce the proton-rich isotopes between $Z=82$ and $Z=100$.
In Ref. \cite{2016Celikovic}, from a fragmentation reaction of the $^{124}$Xe beam impinging on the beryllium target performed in RIKEN, new proton-rich nuclides, $^{96}$In, $^{94}$Cd, $^{92}$Ag, and $^{90}$Pd were observed.
With the fusion-evaporation reaction $^{50}{\rm Cr} + ^{58}{\rm Ni}$ in GANIL \cite{1996Chartier}, the secondary ions of $^{100}$Ag, $^{100}$Cd, $^{100}$In, and $^{100}$Sn were produced.
Recently, the fusion-evaporation reaction $^{54}{\rm Fe} (^{58}{\rm Ni},4n) ^{108}{\rm Xe}$ has reported the first observation of the alpha-decay chain,
$^{108}$Xe$\rightarrow  ^{104}$Te$\rightarrow  ^{100}$Sn \cite{2018Auranen}.
In spite of these experimental developments, no direct evidence has been reported on the binding nature of $^{102}$Sb.

In this work, we theoretically investigate the one-proton emission from $^{102}$Sb, considering an unbound and meta-stable system, in which the proton-energy spectrum should have a finite width corresponding to its lifetime.
For this purpose, the time-dependent three-body model is utilized \cite{2018Oishi_LA, 2018Oishi}.

\section{Model and formalism} \label{Sec:2}
\begin{figure}[t] \begin{center}
\includegraphics[width=\hsize]{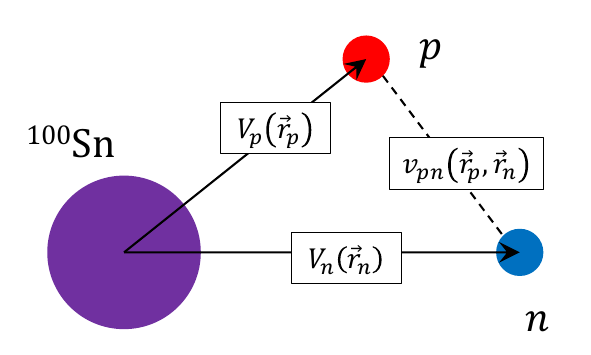}
\caption{Three-body model of $^{102}$Sb.}
\label{fig:chain}
\end{center} \end{figure}

Employing the core-orbital coordinates $\left\{ \bir_p,\bir_n \right\}$ \cite{2016Tani,2012Tani,2016Masui}, we assume the $^{102}$Sb nucleus as the three-body system $^{100}{\rm Sn}+p+n$, which is described in FIG. \ref{fig:chain}.
Our three-body Hamiltonian is given as \cite{2016Tani,2012Tani,2016Masui}
\beq
 H_{3B} = h_p(\bi{r}_p) + h_n(\bi{r}_n) + f_{pn}v_{pn}(\bir_p, \bir_n) +\frac{\bi{p}_p \cdot \bi{p}_n}{m_c}, \label{eq:fevag}
\eeq
where $k=p$ ($k=n$) for the valence proton (neutron).
Here $\bir_k$ indicates the relative coordinate between the core and the $k$-th nucleon,
and thus, $h_p$ ($h_n$) is the core-proton (core-neutron) Hamiltonian.
That is
\beq
 h_k(\bir_k) = \frac{p_k^2}{2\mu_k} + V_{k}(r_k).
\eeq
Mass parameters are given as $\mu_k=m_k m_c/(m_k + m_c)$, where
$m_p=938.272$ MeV$/c^2$, 
$m_n=939.565$ MeV$/c^2$, and
$m_c=50m_p +50m_n -825.16$ MeV$/c^2$ of the $^{100}$Sn core \cite{NNDC_Chart}.

\begin{table}[b]
\begin{center}
\caption{Parameters of the potentials $V_p$ and $V_n$.}\label{table:TAO3}
\begingroup \renewcommand{\arraystretch}{1.2}
\begin{tabular*}{\hsize}{ @{\extracolsep{\fill}} rccccc}
  \hline  \hline
                &$A_C$  &$r_0$    &$a_0$   &$V_0$    &$U_{ls}$~\\
                &       &[fm]    &[fm]    &[MeV]    &[MeV$\cdot$fm$^2$]~ \\  \hline
$^{100}$Sn-$p/n$~&$100$  &$1.24$  &$0.63$   &$-52.1$ &$32.8 r_0^2$~ \\
$^{40}$Ca-$p/n$~&$40$   &$1.25$  &$0.65$   &$-55.7$ &$10.8 r_0^2$~ \\
$^{16}$O-$p/n$~&$16$    &$1.25$  &$0.65$   &$-53.2$ &$22.1 r_0^2$~ \\
  \hline  \hline
\end{tabular*}
\endgroup
\end{center}
\end{table}

The spherical symmetry is assumed for both $V_{p}(r_p)$ and $V_{n}(r_n)$.
These potentials are determined as
\beqa
&& V_{k}(r_k) = V_{WS}(r_k) + V_{Coul}(r_k) \delta_{k,p}, \label{eq:mx6} \\
&& V_{WS}(r) = V_0 f(r) + (\bi{l} \cdot \bi{s}) \frac{U_{ls}}{r} \frac{df(r)}{dr}, \label{eq:cp_WS}
\eeqa
where $R_0=r_0 A_C^{1/3}$ and $f(r)=\left[ 1 + e^{(r-R_0)/a_0} \right]^{-1}$ (Fermi profile).
Parameters are listed in TABLE \ref{table:TAO3}. 
In addition, the Coulomb potential $V_{Coul}(r_p)$ of an uniformly charged sphere with radius $R_0$ is employed for the core-proton subsystem.
\tomo{We mention that the energy of the valence proton is insensitive to the interior profile of the Coulomb potential $V_{Coul}(r_p)$,
as long as the potential barrier around the surface, $r_p \cong r_0 A_C^{1/3}$, remains unchanged.}

\begin{table}[b] \begin{center}
\caption{Single-nucleon energies obtained in this work.
For the resonances in $^{101}$Sb$=^{100}$Sn$+p$, their positive energies are determined by solving the scattering phase shift.
The results of Skyrme-meanfield calculations with SLy4 and SkM* parameters are presented for comparison.
The unit is MeV.
}
\label{table:TAO2}
\catcode`? = \active \def?{\phantom{0}} 
\begingroup \renewcommand{\arraystretch}{1.2}
\begin{tabular*}{\hsize}{ @{\extracolsep{\fill}} crrrr }
\hline \hline
     &          &SLy4 \cite{1997Chabanat_SLY4_v1,1998Chabanat_SLY4_v2}  &SkM* \cite{SKMS}   &This work\\  \hline
$^{100}$Sn-$n$ &$1f_{5/2?}$  &$-22.214$  &$-20.714$   &$-18.715$  \\
      &$2p_{1/2?}$  &$-19.890$  &$-19.090$   &$-17.203$  \\
      &$1g_{9/2?}$  &$-16.948$  &$-16.995$   &$-16.727$  \\
      &$2d_{5/2?}$  &$-10.849$  &$-11.114$   &$-10.822$  \\
      &$1h_{11/2}$  &$?-6.301$  &$?-7.208$   &$?-8.413$  \\
\hline
$^{100}$Sn-$p$ &$1f_{5/2?}$ &$?-7.987$  &$?-6.629$    &$?-6.054$  \\
            &$2p_{1/2?}$ &$?-5.674$  &$?-4.929$    &$?-3.737$  \\
            &$1g_{9/2?}$ &$?-3.081$  &$?-3.150$    &$?-2.576$  \\
            &$2d_{5/2?}$ &$-$        &$-$          &$?+3.414$  \\
            &$1h_{11/2}$ &$-$        &$-$          &$?+5.221$  \\
\hline \hline
\end{tabular*}
\endgroup
\catcode`? = 12 
\end{center} \end{table}

\begin{figure}[t] \begin{center}
 \includegraphics[width=0.9\hsize]{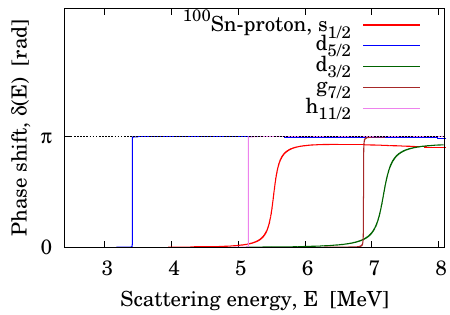}
 \caption{Scattering phase shift calculated with the $^{100}$Sn-proton potential $V_p(r)$.} \label{fig:scat}
\end{center} \end{figure}

In TABLE \ref{table:TAO2}, the single-nucleon energies solved with $V_n$ and $V_p$ are summarized.
Our core-nucleon potentials are tuned so as to reproduce the neutron energies in $2d_{5/2}$ and $1g_{9/2}$ evaluated by the spherical Skyrme-meanfield calculations with the SLy4 parameters \cite{72Vautherin,1997Chabanat_SLY4_v1,1998Chabanat_SLY4_v2}.
Notice that our neutron energy in the $2d_{5/2}$ level is $-10.822$ MeV.
This value is consistent to the experimental one-neutron separation energy,
$S_n = 11.2(4)$ MeV, of the $^{101}$Sn nucleus \cite{NNDC_Chart}.

For the valence proton, levels up to the $1g_{9/2}$ are bound consistently to the magic number $Z=50$ in the core.
For the additional valence proton, no bound orbits are available.
Namely, the $^{101}$Sb nucleus is particle-unbound in the present model.
Instead, several resonances are predicted.
For this purpose, the scattering problem with $h_p(\bir_p)$ is solved, and its phase shifts $\delta(E)$ are displayed in FIG. \ref{fig:scat}.
The lowest $d_{5/2}$ resonance exists at $E_p =3.41$ MeV.
From numerical fitting with
$\delta (E) \cong \arctan \left( \frac{E-E_p}{\Gamma_p /2} \right) +\frac{\pi}{2}$,
its width is obtained as $\Gamma_p \cong 1.5 \times 10^{-4}$ MeV.

We employ the $pn$ interaction of the finite-range Gaussian type.
That is
\beqa
v_{pn}(\bir_p, \bir_n) &=& \left[ V_R e^{-a_R d^2}  +V_S e^{-a_S d^2} \right] \hat{P}_{S=0} \nonumber  \\
&& +\left[ V_R e^{-a_R d^2}  +V_T e^{-a_T d^2} \right] \hat{P}_{S=1}, \label{eq:DIYV}
\eeqa
where $d \equiv \abs{\bir_p -\bir_n}$.
The operator $\hat{P}_{S=0}$ ($\hat{P}_{S=1}$) indicates the projection into the
spin-singlet (spin-triplet) channel of the $pn$ subsystem.
Parameters read $V_R = 200$ MeV, $V_S = -91.85$ MeV, $V_T = -178$ MeV,
$a_R = 1.487$ fm$^{-2}$, $a_S = 0.465$ fm$^{-2}$, and $a_T = 0.639$ fm$^{-2}$ \cite{77Thom}.

The two-body binding energy of isolated deuteron, $E_d$, is solved as the lowest eigenenergy of $H_{d}$,
which reads
\beq
H_{d}(\bi{d}) = \frac{{\bi p}^2_{\bi{d}}}{2\mu_d} +v_{pn}(\bi{d}),
\eeq
where $\bi{d} \equiv \bir_p -\bir_n $ and $\mu_d = m_p m_n/(m_p +m_n)$.
By using $v_{pn}$ in Eq. (\ref{eq:DIYV}), the empirical value is reproduced: $E_d = -2.20$ MeV.
Note that the spin-triplet and $s$-wave state is assumed.
In the following sections, the tuning factor
is also employed to modulate the strength of the $pn$ interaction (deuteron-correlation energy):
\beq
v_{pn}(\bir_p, \bir_n)  \longrightarrow f_{pn} v_{pn}(\bir_p, \bir_n).
\eeq
Thus, $f_{pn}=1$ indicates the bare-deuteron energy.

\begin{figure}[t] \begin{center}
\includegraphics[width=\hsize]{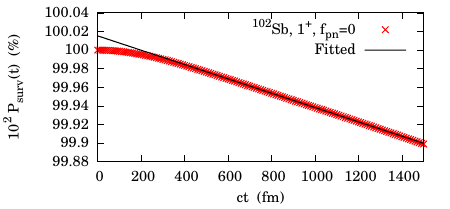}
\caption{Survival probability with $f_{pn}=0$: the $pn$ interaction is neglected.
The fitted function reads $P(t) = P(0) \exp (-t \Gamma_p /\hbar )$ with $P(0)=1.00015$ and $\Gamma_p = 1.5212 \times 10^{-4}$ MeV.
}\label{fig:GGG4}
\end{center} \end{figure}
\begin{figure}[t] \begin{center}
\includegraphics[width=\hsize]{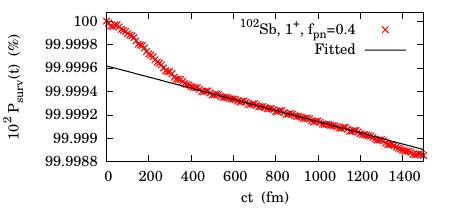}
\caption{Same to FIG. \ref{fig:GGG4} but with $f_{pn}=0.4$.
The fitted function reads $P(t) = P(0) \exp (-t \Gamma_p /\hbar )$ with $P(0)=0.999996$ and $\Gamma_p = 9.3965 \times 10^{-7}$ MeV.
}\label{fig:GGG04}
\end{center} \end{figure}

The three-body Hamiltonian $H_{3B}$ is diagonalized with the basis
$\left[  \phi_p (\bir_p) \otimes \phi_n (\bir_n) \right]^{(J,\pi)}$ for the $J^{\pi}$ state,
where $J$ is the total angular momentum, and $\pi = (-)^{l_p + l_n}$.
Here $\left\{ \phi_k (\bir_k) \right\}$ ($k=p/n$) are the single-nucleon wave functions:
$h_k \ket{\phi_k} = \epsilon_k \ket{\phi_k}$.
The basis up to $24$ MeV are taken into account.
However, because of the Pauli principle, single-proton states up to
the $1g_{9/2}$ for $Z=50$ in the $^{100}$Sn core are excluded.
The same exclusion applies to the neutron states for $N=50$.

\tomo{For the $pn$-coupled spin $\vec{S}_{pn}=\vec{s}_p + \vec{s}_n$,
we mention that this quantity cannot be a well-defined quantum number
in the three-body system, but only the
$\vec{J}= \vec{l}_p + \vec{l}_n + \vec{s}_p + \vec{s}_n$ has well-defined eigenvalues.
Thus, there are both $S_{pn}=0$ and $S_{pn}=1$ components involved in the unique-$J^{\pi}$ state.
The corresponding mixture of the isospin components happens.}

For studying the particle-unbound cases, we employ the time-dependent calculation.
The same calculation was utilized in Ref. \cite{2018Oishi}.
The initial state, $\ket{\Psi (t=0)}$, is solved within the confining Hamiltonian,
\beq
H'_{3B} = H_{3B} +\Delta V_p(r_p),
\eeq
where $\Delta V_p$ is the confining potential for the valence proton.
Note that the valence neutron is bound in this work.

For $t>0$, the time evolution is solved with the original Hamiltonian:
\beq
\ket{\Psi (t)} =\exp \left[ -it \frac{H_{3B}}{\hbar } \right] \ket{\Psi (0)}.
\eeq
With the survival coefficient $\beta(t) = \Braket{ \Psi (0) \mid \Psi(t)}$,
the decaying state is defined as
\beq
\ket{\Psi_{dcy}(t)} = \ket{\Psi (t)} -\beta(t) \ket{\Psi (0)}.
\eeq
Notice that $\Braket{ \Psi (0) \mid \Psi_{dcy} (t)}=0$.
The survival probability, $P(t) = \abs{\beta (t)}^2 = 1-\Braket{\Psi_{dcy} (t) \mid \Psi_{dcy} (t)}$,
can be approximated as the exponential damping if only one resonance exists.

\begin{figure}[t] \begin{center}
\includegraphics[width=0.9\hsize]{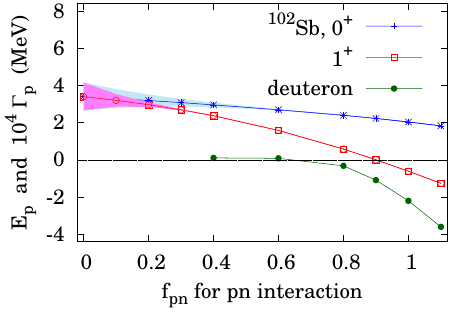}
\caption{One-proton energy $E_{p}$ and width $\Gamma_{p}$ (shaded area) of $^{102}$Sb.
The width is multiplied by $10^4$ for visualization.
\tomo{Notice that the sign convention of $E_p$ is opposite to the binding energy: the negative energy, $E_p <0$, indicates the bound state.}
For $f_{pn} \geq 0.6$, the width is not available due to the evaluation limit.
Binding energy of isolated deuteron with the same interaction is also plotted, where the $f_{pn}=1$ reproduces its experimental value $-2.20$ MeV.}
\label{fig:EGP}
\end{center} \end{figure}

\section{Results and discussions} \label{Sec:3_results}
First we neglect the $pn$ interaction ($f_{pn}=0$).
In this no-interaction limit, the $pn$-state is purely of $(p,d_{5/2}) \otimes (n,d_{5/2})$,
where $(p,d_{5/2})$ is the lowest resonance: see FIG \ref{fig:scat}.
\tomo{The ratio of $S_{pn}=1$, namely the isospin-singlet deuteron-like component, is evaluated as $44$\%.}

In the no-interaction limit,
the one-proton emission is active in the time evolution,
where its energy is solved as $E_p (f_{pn}=0) = 3.41$ MeV.
We confirmed that the survival probability can be well approximated as the exponential damping:
$P(t) \cong P(0) \exp (-t \Gamma_p /\hbar )$ in FIG. \ref{fig:GGG4}.
By fitting $P(t)$, the proton-emitting width is evaluated as $\Gamma_p = 1.52 \times 10^{-4}$ MeV.
Notice that this width coincides with the result obtained from the core-proton scattering phase shift in FIG. \ref{fig:scat}.
This is trivial because the valence neutron makes no changes on the proton's motion.
The corresponding lifetime is $\tau (f_{pn}=0) = \hbar /\Gamma_p \cong 4.4 \times 10^{-18}$ seconds.

Next we activate the $pn$ interaction.
The results are presented in FIG \ref{fig:EGP}.
The potential $v_{pn}$ is adjusted to reproduce the binding energy of deuteron when $f_{pn}=1$.
In this case, the $J^{\pi}=1^+$ state of $^{102}$Sb becomes bound against the proton emission.
The additional attraction $v_{pn}$ reduces the valence proton's energy below the threshold.
That is
\beq
E_p = \braket{H_{3B}(f_{pn}=1)} +S_n(^{101}{\rm Sn}) = -0.61~{\rm MeV},
\eeq
where $S_n(^{101}{\rm Sn}) =10.82$ MeV.
\tomo{Note that the sign convention of $E_p$ is opposite to the empirical binding energy: negative $E_p$ values mean that the proton is bound.}
We also checked that, in the time-dependent calculation, no outgoing components can be observed.

\begin{table}[b] \begin{center}
\caption{Properties of the $1^{+}$ state in $^{102}$Sb calculated by changing the $pn$-interaction strength.
The width $\Gamma_p$ with $0.4 < f_{pn}$ is beyond the evaluation limit.
With $f_{pn}=1$, the proton is bound.}
\label{table:FOCUS}
\catcode`? = \active \def?{\phantom{0}} 
\begingroup \renewcommand{\arraystretch}{1.2}
\begin{tabular*}{\hsize}{ @{\extracolsep{\fill}} lrrrr}
\hline \hline
~$f_{pn}$                       &$0$    &$0.4$  &$0.8$~  &$1.0$~\\  \hline
~$\Braket{H_{3B}}$~[MeV]        &$-7.41$  &$-8.44$  &$-10.24$~ &$-11.43$~\\
~$\Braket{f_{pn} v_{pn}}$~[MeV]  &$0$        &$-1.37$  &$-4.43$~ &$-6.34$~\\
~$(p,d_{5/2})\otimes (n,d_{5/2})$~[\%]  &$99.9$  &$93.6$  &$71.7$~  &$60.4$~\\
~$\Gamma_{p}$~[eV]  &$152$ &$\cong 0.94 $               &($\leq 0.1$)  & $0$  \\
\hline \hline
\end{tabular*}
\endgroup
\catcode`? = 12 
\end{center} \end{table}

In FIG \ref{fig:EGP}, the one-proton energy and width are plotted as functions of $f_{pn}$ for $1^+$ and $0^+$ configurations.
In the no-interaction limit, the two resonances are identical, where the single-proton emission of $d_{5/2}$ is allowed.
By increasing $f_{pn}$, the $1^+$ ($0^+$) energy rapidly (slowly) decreases.
Then at the experimental point of deuteron,
the $1^+$ becomes bound, whereas the $0^+$ is still unbound.
The threshold for binding is found at $f_{pn} \geq 0.92$ in the present model.

In TABLE \ref{table:FOCUS}, the three-body energy $\Braket{H_{3B}}$,
the mean $pn$~interaction $\Braket{f_{pn} v_{pn}}$, and
the ratio of $(p,d_{5/2}) \otimes (n,d_{5/2})$ are summarized for several $f_{pn}$ values.
One can read that the $1^{+}$ state becomes more deeply bound with the stronger $pn$ interaction.
In correspondence, the smaller $(p,d_{5/2})\otimes (n,d_{5/2})$ ratio is obtained.
This is because, for deeper binding, an inclusion of continuum states other than $(p,d_{5/2})$ is necessary.
The same effect is known for stabilizing two-neutron Borromean nuclei \cite{1991BE, 1997EBH, 2005HS, 07Bertulani_76}.

\begin{table}[b] \begin{center}
\caption{The $pn$-separation energies, $S_{pn,{\rm calc.}} = -\Braket{H_{3B}}$, calculated in this work.
Experimental data of $^{42}$Sc and $^{18}$F are presented for comparison.
Note that $^{42}$Sc~($1^+$) is not the ground but the first-excited state. 
}
\label{table:TAO87}
\catcode`! = \active \def!{\phantom{0}} 
\begingroup \renewcommand{\arraystretch}{1.2}
\begin{tabular*}{\hsize}{ @{\extracolsep{\fill}} rcccc }
\hline \hline
          &$f_{pn}$  &$S_{pn,{\rm calc.}}$      &$S_{pn,{\rm expt.}}$~\cite{NNDC_Chart}   &$p$-emitter~?~\\
          &         &[MeV]    &[MeV]    &~ \\  \hline
$^{102}$Sb~($1^+$)~&$1.00$   &$11.43$    &$-$ &No~ \\
                 &$0.60$    &$!9.23$    &$-$ &Yes~ \\
                 &$0.40$    &$!8.44$    &$-$ &Yes~ \\
 $^{42}$Sc~($1^+$)~&$0.38$   &$12.06$    &$12.023$  &No~ \\
  $^{18}$F~($1^+$)~&$0.58$   &$!9.75$    &$!9.750$  &No~ \\
\hline \hline
\end{tabular*}
\endgroup
\catcode`! = 12 
\end{center} \end{table}

\begin{figure}[t] \begin{center}
\includegraphics[width=\hsize]{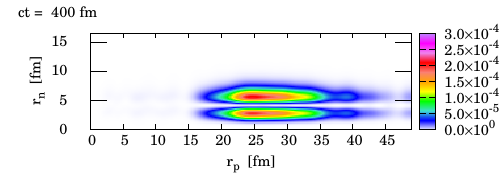}
\includegraphics[width=\hsize]{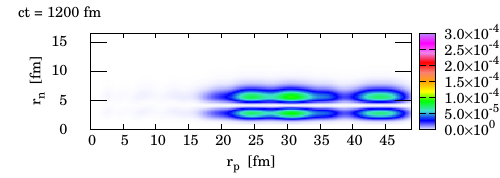}
\caption{Time-dependent density distribution of the decaying state, $\rho_{dcy} (t,r_p,r_n)$, with $f_{pn}=0.4$.
}\label{fig:GGG2}
\end{center} \end{figure}

For computing finite nuclei, one often needs to weaken the $pn$ interaction in order to reproduce the empirical energies.
\tomo{For the valence proton and neutron around the nuclear surface,
because of the density dependence, the effective $pn$ interaction can become reduced \cite{1968Tamagaki,1970Tamagaki,1982Suzuki_NPA,1982Suzuki_PTP}.
For inferring this weakening effect phenomenologically, we performed the same three-body calculations for
$^{18}$F ($^{16}{\rm O}+p+n$) and
$^{42}$Sc ($^{40}{\rm Ca}+p+n$) nuclei.}
For their core-nucleon potentials, parameters are given in TABLE \ref{table:TAO3}.
These parameters are optimized so as to reproduce the experimental single-nucleon energies \cite{2019OP_M1}.
For the $pn$ interaction, the same finite-range Gaussian potential in Eq. (\ref{eq:DIYV}) is utilized.
The results are presented in TABLE \ref{table:TAO87}.
Consequently, a finite weakening with $f_{pn}=0.38$ ($f_{pn}=0.58$) is necessary to
reproduce the $pn$-separation energy of the $^{42}$Sc ($^{18}$F) nucleus.
By assuming that the same weakening is necessary in $^{102}$Sb, the proton emission is expected from its $1^+$ state.

As an example, the results with $f_{pn} = 0.4$ are displayed in FIGs. \ref{fig:GGG04} and \ref{fig:GGG2}:
this case corresponds to the limit for quantitatively evaluating $\Gamma_{p}$ with the present model.
In FIG. \ref{fig:GGG04}, the survival probability is plotted.
The proton-emission width of $^{102}$Sb ($1^+$) is evaluated as $\Gamma_p \cong 9.4 \times 10^{-7}$ MeV, where the corresponding lifetime reads $7.0 \times 10^{-16}$ seconds.
Notice that the polynomial behavior, $P(t) \cong 1 - a (t/ \tau)^2$, is observed in the early-time region.
Such a behavior is generally known in time development, being relevant to
the quantum Zeno effect \cite{1977Misra,1977Chiu,1988Levitan,2021Jimenez}.

In FIG. \ref{fig:GGG2}, the normalized density distribution of the time-dependent decaying state
is displayed:
\beqa
\rho_{dcy} (t,r_p,r_n) &=& \frac{r^2_p r^2_n }{N_{dcy}(t)} \iint d\Omega_p d\Omega_n  \nonumber  \\
&& \abs{\Psi_{dcy} (t, \bir_p, \bir_n)}^2,
\eeqa
where $N_{dcy} (t) = \Braket{\Psi_{dcy} (t) \mid \Psi_{dcy} (t)}$.
Its angular components are integrated, and thus, $\iint dr_p dr_n \rho_{dcy} (t,r_p,r_n) =1$.
One can observe the trajectory in the region with $r_p >10$ fm and $r_n \leq 10$ fm.
It can be interpreted as that the unbound proton is emitted outside, whereas the bound neutron stays around the core.

With $f_{pn} > 0.4$, one can expect the narrower width as well as the longer lifetime.
In such cases with extremely narrow widths, however, the quantitative evaluation is still demanding. 

\section{Summary} \label{Sec:4}
The $^{102}$Sb nucleus is suggested as a proton emitter.
We reached this conclusion by considering the weakening of $pn$ interaction.
This weakening effect is introduced from the $pn$-separation energies of other core-$pn$ systems, $^{42}$Sc and $^{18}$F.
In this proton-emitting scenario,
the lower limit of lifetime is evaluated as $\tau \gtrsim 4.4 \times 10^{-18}$ seconds in both the $1^+$ and $0^+$ cases.
This limit is determined by assuming no interactions between the valence proton and neutron.
However, since the finite $pn$ interaction actually exists, the longer lifetime is expected.
\tomo{Note that the $pn$ interaction and its weakening effect have been treated phenomenologically in this work.
Further sophistication of models are expected in forthcoming studies.}

In this work, the valence neutron is bound.
From excited nuclei, in contrast, the emission of both proton and neutron can take place.
Especially the deuteron emission and its sensitivity to the $pn$ interaction is left for future works.

\begin{acknowledgments}
T. Oishi appreciate the suggestions by Akira Ohnishi, Masaki Sasano, and Hiroyuki Sagawa.
Numerical calculations are supported by the cooperative project of supercomputer Yukawa-21 in Yukawa Institute for Theoretical Physics, Kyoto University.
T. Oishi acknowledge the Multi-disciplinary Cooperative Research Program (MCRP) in FY2023 and FY2024 (project ID wo23i034) by Center for Computational Sciences, University of Tsukuba, allocating computational resources of supercomputer Wisteria/BDEC-01 (Odyssey) in Information Technology Center, University of Tokyo.
\end{acknowledgments}

\input{SEBNAP.bbl}


\end{document}

%% file: SEBNAP.bbl
%

%% file: SEBNAP.bbl
\begin{thebibliography}{38}%
\makeatletter
\providecommand \@ifxundefined [1]{%
 \@ifx{#1\undefined}
}%
\providecommand \@ifnum [1]{%
 \ifnum #1\expandafter \@firstoftwo
 \else \expandafter \@secondoftwo
 \fi
}%
\providecommand \@ifx [1]{%
 \ifx #1\expandafter \@firstoftwo
 \else \expandafter \@secondoftwo
 \fi
}%
\providecommand \natexlab [1]{#1}%
\providecommand \enquote  [1]{``#1''}%
\providecommand \bibnamefont  [1]{#1}%
\providecommand \bibfnamefont [1]{#1}%
\providecommand \citenamefont [1]{#1}%
\providecommand \href@noop [0]{\@secondoftwo}%
\providecommand \href [0]{\begingroup \@sanitize@url \@href}%
\providecommand \@href[1]{\@@startlink{#1}\@@href}%
\providecommand \@@href[1]{\endgroup#1\@@endlink}%
\providecommand \@sanitize@url [0]{\catcode `\\12\catcode `\$12\catcode
  `\&12\catcode `\#12\catcode `\^12\catcode `\_12\catcode `\%12\relax}%
\providecommand \@@startlink[1]{}%
\providecommand \@@endlink[0]{}%
\providecommand \url  [0]{\begingroup\@sanitize@url \@url }%
\providecommand \@url [1]{\endgroup\@href {#1}{\urlprefix }}%
\providecommand \urlprefix  [0]{URL }%
\providecommand \Eprint [0]{\href }%
\providecommand \doibase [0]{https://doi.org/}%
\providecommand \selectlanguage [0]{\@gobble}%
\providecommand \bibinfo  [0]{\@secondoftwo}%
\providecommand \bibfield  [0]{\@secondoftwo}%
\providecommand \translation [1]{[#1]}%
\providecommand \BibitemOpen [0]{}%
\providecommand \bibitemStop [0]{}%
\providecommand \bibitemNoStop [0]{.\EOS\space}%
\providecommand \EOS [0]{\spacefactor3000\relax}%
\providecommand \BibitemShut  [1]{\csname bibitem#1\endcsname}%
\let\auto@bib@innerbib\@empty
\bibitem [{\citenamefont {Tanimura}\ and\ \citenamefont
  {Sagawa}(2016)}]{2016Tani}%
  \BibitemOpen
  \bibfield  {author} {\bibinfo {author} {\bibfnamefont {Y.}~\bibnamefont
  {Tanimura}}\ and\ \bibinfo {author} {\bibfnamefont {H.}~\bibnamefont
  {Sagawa}},\ }\href {https://doi.org/10.1103/PhysRevC.93.064319} {\bibfield
  {journal} {\bibinfo  {journal} {Phys. Rev. C}\ }\textbf {\bibinfo {volume}
  {93}},\ \bibinfo {pages} {064319} (\bibinfo {year} {2016})}\BibitemShut
  {NoStop}%
\bibitem [{\citenamefont {Kalandarov}\ \emph {et~al.}(2014)\citenamefont
  {Kalandarov}, \citenamefont {Adamian}, \citenamefont {Antonenko},\ and\
  \citenamefont {Wieleczko}}]{2014Kalanda}%
  \BibitemOpen
  \bibfield  {author} {\bibinfo {author} {\bibfnamefont {S.~A.}\ \bibnamefont
  {Kalandarov}}, \bibinfo {author} {\bibfnamefont {G.~G.}\ \bibnamefont
  {Adamian}}, \bibinfo {author} {\bibfnamefont {N.~V.}\ \bibnamefont
  {Antonenko}},\ and\ \bibinfo {author} {\bibfnamefont {J.~P.}\ \bibnamefont
  {Wieleczko}},\ }\href {https://doi.org/10.1103/PhysRevC.90.024609} {\bibfield
   {journal} {\bibinfo  {journal} {Phys. Rev. C}\ }\textbf {\bibinfo {volume}
  {90}},\ \bibinfo {pages} {024609} (\bibinfo {year} {2014})}\BibitemShut
  {NoStop}%
\bibitem [{\citenamefont {Doornenbal}\ \emph {et~al.}(2014)\citenamefont
  {Doornenbal}, \citenamefont {Takeuchi}, \citenamefont {Aoi}, \citenamefont
  {Matsushita}, \citenamefont {Obertelli}, \citenamefont {Steppenbeck},
  \citenamefont {Wang}, \citenamefont {Audirac}, \citenamefont {Baba},
  \citenamefont {Bednarczyk}, \citenamefont {Boissinot}, \citenamefont
  {Ciemala}, \citenamefont {Corsi}, \citenamefont {Furumoto}, \citenamefont
  {Isobe}, \citenamefont {Jungclaus}, \citenamefont {Lapoux}, \citenamefont
  {Lee}, \citenamefont {Matsui}, \citenamefont {Motobayashi}, \citenamefont
  {Nishimura}, \citenamefont {Ota}, \citenamefont {Pollacco}, \citenamefont
  {Sakurai}, \citenamefont {Santamaria}, \citenamefont {Shiga}, \citenamefont
  {Sohler},\ and\ \citenamefont {Taniuchi}}]{2014Doornenbal}%
  \BibitemOpen
  \bibfield  {author} {\bibinfo {author} {\bibfnamefont {P.}~\bibnamefont
  {Doornenbal}}, \bibinfo {author} {\bibfnamefont {S.}~\bibnamefont
  {Takeuchi}}, \bibinfo {author} {\bibfnamefont {N.}~\bibnamefont {Aoi}},
  \bibinfo {author} {\bibfnamefont {M.}~\bibnamefont {Matsushita}}, \bibinfo
  {author} {\bibfnamefont {A.}~\bibnamefont {Obertelli}}, \bibinfo {author}
  {\bibfnamefont {D.}~\bibnamefont {Steppenbeck}}, \bibinfo {author}
  {\bibfnamefont {H.}~\bibnamefont {Wang}}, \bibinfo {author} {\bibfnamefont
  {L.}~\bibnamefont {Audirac}}, \bibinfo {author} {\bibfnamefont
  {H.}~\bibnamefont {Baba}}, \bibinfo {author} {\bibfnamefont {P.}~\bibnamefont
  {Bednarczyk}}, \bibinfo {author} {\bibfnamefont {S.}~\bibnamefont
  {Boissinot}}, \bibinfo {author} {\bibfnamefont {M.}~\bibnamefont {Ciemala}},
  \bibinfo {author} {\bibfnamefont {A.}~\bibnamefont {Corsi}}, \bibinfo
  {author} {\bibfnamefont {T.}~\bibnamefont {Furumoto}}, \bibinfo {author}
  {\bibfnamefont {T.}~\bibnamefont {Isobe}}, \bibinfo {author} {\bibfnamefont
  {A.}~\bibnamefont {Jungclaus}}, \bibinfo {author} {\bibfnamefont
  {V.}~\bibnamefont {Lapoux}}, \bibinfo {author} {\bibfnamefont
  {J.}~\bibnamefont {Lee}}, \bibinfo {author} {\bibfnamefont {K.}~\bibnamefont
  {Matsui}}, \bibinfo {author} {\bibfnamefont {T.}~\bibnamefont {Motobayashi}},
  \bibinfo {author} {\bibfnamefont {D.}~\bibnamefont {Nishimura}}, \bibinfo
  {author} {\bibfnamefont {S.}~\bibnamefont {Ota}}, \bibinfo {author}
  {\bibfnamefont {E.~C.}\ \bibnamefont {Pollacco}}, \bibinfo {author}
  {\bibfnamefont {H.}~\bibnamefont {Sakurai}}, \bibinfo {author} {\bibfnamefont
  {C.}~\bibnamefont {Santamaria}}, \bibinfo {author} {\bibfnamefont
  {Y.}~\bibnamefont {Shiga}}, \bibinfo {author} {\bibfnamefont
  {D.}~\bibnamefont {Sohler}},\ and\ \bibinfo {author} {\bibfnamefont
  {R.}~\bibnamefont {Taniuchi}},\ }\href
  {https://doi.org/10.1103/PhysRevC.90.061302} {\bibfield  {journal} {\bibinfo
  {journal} {Phys. Rev. C}\ }\textbf {\bibinfo {volume} {90}},\ \bibinfo
  {pages} {061302} (\bibinfo {year} {2014})}\BibitemShut {NoStop}%
\bibitem [{\citenamefont {Maharana}\ \emph {et~al.}(2015)\citenamefont
  {Maharana}, \citenamefont {Bhagwat},\ and\ \citenamefont
  {Gambhir}}]{2015Maharana}%
  \BibitemOpen
  \bibfield  {author} {\bibinfo {author} {\bibfnamefont {J.~P.}\ \bibnamefont
  {Maharana}}, \bibinfo {author} {\bibfnamefont {A.}~\bibnamefont {Bhagwat}},\
  and\ \bibinfo {author} {\bibfnamefont {Y.~K.}\ \bibnamefont {Gambhir}},\
  }\href {https://doi.org/10.1103/PhysRevC.91.047301} {\bibfield  {journal}
  {\bibinfo  {journal} {Phys. Rev. C}\ }\textbf {\bibinfo {volume} {91}},\
  \bibinfo {pages} {047301} (\bibinfo {year} {2015})}\BibitemShut {NoStop}%
\bibitem [{\citenamefont {Xu}\ \emph {et~al.}(2019)\citenamefont {Xu},
  \citenamefont {Zhang}, \citenamefont {Li}, \citenamefont {Li}, \citenamefont
  {Sokhna}, \citenamefont {Zhang}, \citenamefont {Yang}, \citenamefont {Cheng},
  \citenamefont {Zhang}, \citenamefont {Ge}, \citenamefont {Li}, \citenamefont
  {Liu},\ and\ \citenamefont {Zhang}}]{2019Xu}%
  \BibitemOpen
  \bibfield  {author} {\bibinfo {author} {\bibfnamefont {X.-X.}\ \bibnamefont
  {Xu}}, \bibinfo {author} {\bibfnamefont {G.}~\bibnamefont {Zhang}}, \bibinfo
  {author} {\bibfnamefont {J.-J.}\ \bibnamefont {Li}}, \bibinfo {author}
  {\bibfnamefont {B.}~\bibnamefont {Li}}, \bibinfo {author} {\bibfnamefont
  {C.~A.~T.}\ \bibnamefont {Sokhna}}, \bibinfo {author} {\bibfnamefont {X.-R.}\
  \bibnamefont {Zhang}}, \bibinfo {author} {\bibfnamefont {X.-X.}\ \bibnamefont
  {Yang}}, \bibinfo {author} {\bibfnamefont {S.-H.}\ \bibnamefont {Cheng}},
  \bibinfo {author} {\bibfnamefont {Y.-H.}\ \bibnamefont {Zhang}}, \bibinfo
  {author} {\bibfnamefont {Z.-S.}\ \bibnamefont {Ge}}, \bibinfo {author}
  {\bibfnamefont {C.}~\bibnamefont {Li}}, \bibinfo {author} {\bibfnamefont
  {Z.}~\bibnamefont {Liu}},\ and\ \bibinfo {author} {\bibfnamefont {F.-S.}\
  \bibnamefont {Zhang}},\ }\href
  {https://doi.org/10.1088/1674-1137/43/6/064105} {\bibfield  {journal}
  {\bibinfo  {journal} {Chinese Physics C}\ }\textbf {\bibinfo {volume} {43}},\
  \bibinfo {pages} {064105} (\bibinfo {year} {2019})}\BibitemShut {NoStop}%
\bibitem [{\citenamefont {Brookhaven National~Laboratory}(2022)}]{NNDC_Chart}%
  \BibitemOpen
  \bibfield  {author} {\bibinfo {author} {\bibfnamefont {N.~N. D.~C.}\
  \bibnamefont {Brookhaven National~Laboratory}},\ }\href
  {https://www.nndc.bnl.gov/nudat3/} {\bibinfo {title} {Chart of nuclides in
  nudat 3.0}} (\bibinfo {year} {2022}),\ \bibinfo {note}
  {https://www.nndc.bnl.gov/nudat3/}\BibitemShut {NoStop}%
\bibitem [{\citenamefont {Tanimura}\ \emph {et~al.}(2012)\citenamefont
  {Tanimura}, \citenamefont {Hagino},\ and\ \citenamefont {Sagawa}}]{2012Tani}%
  \BibitemOpen
  \bibfield  {author} {\bibinfo {author} {\bibfnamefont {Y.}~\bibnamefont
  {Tanimura}}, \bibinfo {author} {\bibfnamefont {K.}~\bibnamefont {Hagino}},\
  and\ \bibinfo {author} {\bibfnamefont {H.}~\bibnamefont {Sagawa}},\ }\href
  {https://doi.org/10.1103/PhysRevC.86.044331} {\bibfield  {journal} {\bibinfo
  {journal} {Phys. Rev. C}\ }\textbf {\bibinfo {volume} {86}},\ \bibinfo
  {pages} {044331} (\bibinfo {year} {2012})}\BibitemShut {NoStop}%
\bibitem [{\citenamefont {Masui}\ and\ \citenamefont
  {Kimura}(2016)}]{2016Masui}%
  \BibitemOpen
  \bibfield  {author} {\bibinfo {author} {\bibfnamefont {H.}~\bibnamefont
  {Masui}}\ and\ \bibinfo {author} {\bibfnamefont {M.}~\bibnamefont {Kimura}},\
  }\href {https://doi.org/10.1093/ptep/ptw041} {\bibfield  {journal} {\bibinfo
  {journal} {Progress of Theoretical and Experimental Physics}\ }\textbf
  {\bibinfo {volume} {2016}},\ \bibinfo {pages} {053D01} (\bibinfo {year}
  {2016})}\BibitemShut {NoStop}%
\bibitem [{\citenamefont {Uzawa}\ \emph {et~al.}(2024)\citenamefont {Uzawa},
  \citenamefont {Hinohara},\ and\ \citenamefont {Nakatsukasa}}]{2024Uzawa}%
  \BibitemOpen
  \bibfield  {author} {\bibinfo {author} {\bibfnamefont {K.}~\bibnamefont
  {Uzawa}}, \bibinfo {author} {\bibfnamefont {N.}~\bibnamefont {Hinohara}},\
  and\ \bibinfo {author} {\bibfnamefont {T.}~\bibnamefont {Nakatsukasa}},\
  }\href {https://doi.org/10.1093/ptep/ptae072} {\bibfield  {journal} {\bibinfo
   {journal} {Progress of Theoretical and Experimental Physics}\ }\textbf
  {\bibinfo {volume} {2024}},\ \bibinfo {pages} {053D02} (\bibinfo {year}
  {2024})}\BibitemShut {NoStop}%
\bibitem [{\citenamefont {Devaraja}\ \emph {et~al.}(2015)\citenamefont
  {Devaraja}, \citenamefont {Heinz}, \citenamefont {Beliuskina}, \citenamefont
  {Comas}, \citenamefont {Hofmann}, \citenamefont {Hornung}, \citenamefont
  {M\"{u}nzenberg}, \citenamefont {Nishio}, \citenamefont {Ackermann},
  \citenamefont {Gambhir}, \citenamefont {Gupta}, \citenamefont {Henderson},
  \citenamefont {He{\ss}berger}, \citenamefont {Khuyagbaatar}, \citenamefont
  {Kindler}, \citenamefont {Lommel}, \citenamefont {Moody}, \citenamefont
  {Maurer}, \citenamefont {Mann}, \citenamefont {Popeko}, \citenamefont
  {Shaughnessy}, \citenamefont {Stoyer},\ and\ \citenamefont
  {Yeremin}}]{2015Devaraja}%
  \BibitemOpen
  \bibfield  {author} {\bibinfo {author} {\bibfnamefont {H.}~\bibnamefont
  {Devaraja}}, \bibinfo {author} {\bibfnamefont {S.}~\bibnamefont {Heinz}},
  \bibinfo {author} {\bibfnamefont {O.}~\bibnamefont {Beliuskina}}, \bibinfo
  {author} {\bibfnamefont {V.}~\bibnamefont {Comas}}, \bibinfo {author}
  {\bibfnamefont {S.}~\bibnamefont {Hofmann}}, \bibinfo {author} {\bibfnamefont
  {C.}~\bibnamefont {Hornung}}, \bibinfo {author} {\bibfnamefont
  {G.}~\bibnamefont {M\"{u}nzenberg}}, \bibinfo {author} {\bibfnamefont
  {K.}~\bibnamefont {Nishio}}, \bibinfo {author} {\bibfnamefont
  {D.}~\bibnamefont {Ackermann}}, \bibinfo {author} {\bibfnamefont
  {Y.}~\bibnamefont {Gambhir}}, \bibinfo {author} {\bibfnamefont
  {M.}~\bibnamefont {Gupta}}, \bibinfo {author} {\bibfnamefont
  {R.}~\bibnamefont {Henderson}}, \bibinfo {author} {\bibfnamefont
  {F.}~\bibnamefont {He{\ss}berger}}, \bibinfo {author} {\bibfnamefont
  {J.}~\bibnamefont {Khuyagbaatar}}, \bibinfo {author} {\bibfnamefont
  {B.}~\bibnamefont {Kindler}}, \bibinfo {author} {\bibfnamefont
  {B.}~\bibnamefont {Lommel}}, \bibinfo {author} {\bibfnamefont
  {K.}~\bibnamefont {Moody}}, \bibinfo {author} {\bibfnamefont
  {J.}~\bibnamefont {Maurer}}, \bibinfo {author} {\bibfnamefont
  {R.}~\bibnamefont {Mann}}, \bibinfo {author} {\bibfnamefont {A.}~\bibnamefont
  {Popeko}}, \bibinfo {author} {\bibfnamefont {D.}~\bibnamefont {Shaughnessy}},
  \bibinfo {author} {\bibfnamefont {M.}~\bibnamefont {Stoyer}},\ and\ \bibinfo
  {author} {\bibfnamefont {A.}~\bibnamefont {Yeremin}},\ }\href
  {https://doi.org/https://doi.org/10.1016/j.physletb.2015.07.006} {\bibfield
  {journal} {\bibinfo  {journal} {Physics Letters B}\ }\textbf {\bibinfo
  {volume} {748}},\ \bibinfo {pages} {199} (\bibinfo {year}
  {2015})}\BibitemShut {NoStop}%
\bibitem [{\citenamefont {Kalandarov}\ \emph {et~al.}(2020)\citenamefont
  {Kalandarov}, \citenamefont {Adamian}, \citenamefont {Antonenko},
  \citenamefont {Devaraja},\ and\ \citenamefont {Heinz}}]{2020Kalandarov}%
  \BibitemOpen
  \bibfield  {author} {\bibinfo {author} {\bibfnamefont {S.~A.}\ \bibnamefont
  {Kalandarov}}, \bibinfo {author} {\bibfnamefont {G.~G.}\ \bibnamefont
  {Adamian}}, \bibinfo {author} {\bibfnamefont {N.~V.}\ \bibnamefont
  {Antonenko}}, \bibinfo {author} {\bibfnamefont {H.~M.}\ \bibnamefont
  {Devaraja}},\ and\ \bibinfo {author} {\bibfnamefont {S.}~\bibnamefont
  {Heinz}},\ }\href {https://doi.org/10.1103/PhysRevC.102.024612} {\bibfield
  {journal} {\bibinfo  {journal} {Phys. Rev. C}\ }\textbf {\bibinfo {volume}
  {102}},\ \bibinfo {pages} {024612} (\bibinfo {year} {2020})}\BibitemShut
  {NoStop}%
\bibitem [{\citenamefont {Hencheck}\ \emph {et~al.}(1994)\citenamefont
  {Hencheck}, \citenamefont {Boyd}, \citenamefont {Hellstr\"om}, \citenamefont
  {Morrissey}, \citenamefont {Balbes}, \citenamefont {Chloupek}, \citenamefont
  {Fauerbach}, \citenamefont {Mitchell}, \citenamefont {Pfaff}, \citenamefont
  {Powell}, \citenamefont {Raimann}, \citenamefont {Sherrill}, \citenamefont
  {Steiner}, \citenamefont {Vandegriff},\ and\ \citenamefont
  {Yennello}}]{1994Hencheck}%
  \BibitemOpen
  \bibfield  {author} {\bibinfo {author} {\bibfnamefont {M.}~\bibnamefont
  {Hencheck}}, \bibinfo {author} {\bibfnamefont {R.~N.}\ \bibnamefont {Boyd}},
  \bibinfo {author} {\bibfnamefont {M.}~\bibnamefont {Hellstr\"om}}, \bibinfo
  {author} {\bibfnamefont {D.~J.}\ \bibnamefont {Morrissey}}, \bibinfo {author}
  {\bibfnamefont {M.~J.}\ \bibnamefont {Balbes}}, \bibinfo {author}
  {\bibfnamefont {F.~R.}\ \bibnamefont {Chloupek}}, \bibinfo {author}
  {\bibfnamefont {M.}~\bibnamefont {Fauerbach}}, \bibinfo {author}
  {\bibfnamefont {C.~A.}\ \bibnamefont {Mitchell}}, \bibinfo {author}
  {\bibfnamefont {R.}~\bibnamefont {Pfaff}}, \bibinfo {author} {\bibfnamefont
  {C.~F.}\ \bibnamefont {Powell}}, \bibinfo {author} {\bibfnamefont
  {G.}~\bibnamefont {Raimann}}, \bibinfo {author} {\bibfnamefont {B.~M.}\
  \bibnamefont {Sherrill}}, \bibinfo {author} {\bibfnamefont {M.}~\bibnamefont
  {Steiner}}, \bibinfo {author} {\bibfnamefont {J.}~\bibnamefont
  {Vandegriff}},\ and\ \bibinfo {author} {\bibfnamefont {S.~J.}\ \bibnamefont
  {Yennello}},\ }\href {https://doi.org/10.1103/PhysRevC.50.2219} {\bibfield
  {journal} {\bibinfo  {journal} {Phys. Rev. C}\ }\textbf {\bibinfo {volume}
  {50}},\ \bibinfo {pages} {2219} (\bibinfo {year} {1994})}\BibitemShut
  {NoStop}%
\bibitem [{\citenamefont {Rykaczewski}\ \emph {et~al.}(1995)\citenamefont
  {Rykaczewski}, \citenamefont {Anne}, \citenamefont {Auger}, \citenamefont
  {Bazin}, \citenamefont {Borcea}, \citenamefont {Borrel}, \citenamefont
  {Corre}, \citenamefont {D\"orfler}, \citenamefont {Fomichov}, \citenamefont
  {Grzywacz}, \citenamefont {Guillemaud-Mueller}, \citenamefont {Hue},
  \citenamefont {Huyse}, \citenamefont {Janas}, \citenamefont {Keller},
  \citenamefont {Lewitowicz}, \citenamefont {Lukyanov}, \citenamefont
  {Mueller}, \citenamefont {Penionzhkevich}, \citenamefont {Pf\"utzner},
  \citenamefont {Pougheon}, \citenamefont {Saint-Laurent}, \citenamefont
  {Schmidt}, \citenamefont {Schmidt-Ott}, \citenamefont {Sorlin}, \citenamefont
  {Szerypo}, \citenamefont {Tarasov}, \citenamefont {Wauters},\ and\
  \citenamefont {\ifmmode~\dot{Z}\else \.{Z}\fi{}ylicz}}]{1995Rykacz}%
  \BibitemOpen
  \bibfield  {author} {\bibinfo {author} {\bibfnamefont {K.}~\bibnamefont
  {Rykaczewski}}, \bibinfo {author} {\bibfnamefont {R.}~\bibnamefont {Anne}},
  \bibinfo {author} {\bibfnamefont {G.}~\bibnamefont {Auger}}, \bibinfo
  {author} {\bibfnamefont {D.}~\bibnamefont {Bazin}}, \bibinfo {author}
  {\bibfnamefont {C.}~\bibnamefont {Borcea}}, \bibinfo {author} {\bibfnamefont
  {V.}~\bibnamefont {Borrel}}, \bibinfo {author} {\bibfnamefont {J.~M.}\
  \bibnamefont {Corre}}, \bibinfo {author} {\bibfnamefont {T.}~\bibnamefont
  {D\"orfler}}, \bibinfo {author} {\bibfnamefont {A.}~\bibnamefont {Fomichov}},
  \bibinfo {author} {\bibfnamefont {R.}~\bibnamefont {Grzywacz}}, \bibinfo
  {author} {\bibfnamefont {D.}~\bibnamefont {Guillemaud-Mueller}}, \bibinfo
  {author} {\bibfnamefont {R.}~\bibnamefont {Hue}}, \bibinfo {author}
  {\bibfnamefont {M.}~\bibnamefont {Huyse}}, \bibinfo {author} {\bibfnamefont
  {Z.}~\bibnamefont {Janas}}, \bibinfo {author} {\bibfnamefont
  {H.}~\bibnamefont {Keller}}, \bibinfo {author} {\bibfnamefont
  {M.}~\bibnamefont {Lewitowicz}}, \bibinfo {author} {\bibfnamefont
  {S.}~\bibnamefont {Lukyanov}}, \bibinfo {author} {\bibfnamefont {A.~C.}\
  \bibnamefont {Mueller}}, \bibinfo {author} {\bibfnamefont {Y.}~\bibnamefont
  {Penionzhkevich}}, \bibinfo {author} {\bibfnamefont {M.}~\bibnamefont
  {Pf\"utzner}}, \bibinfo {author} {\bibfnamefont {F.}~\bibnamefont
  {Pougheon}}, \bibinfo {author} {\bibfnamefont {M.~G.}\ \bibnamefont
  {Saint-Laurent}}, \bibinfo {author} {\bibfnamefont {K.}~\bibnamefont
  {Schmidt}}, \bibinfo {author} {\bibfnamefont {W.~D.}\ \bibnamefont
  {Schmidt-Ott}}, \bibinfo {author} {\bibfnamefont {O.}~\bibnamefont {Sorlin}},
  \bibinfo {author} {\bibfnamefont {J.}~\bibnamefont {Szerypo}}, \bibinfo
  {author} {\bibfnamefont {O.}~\bibnamefont {Tarasov}}, \bibinfo {author}
  {\bibfnamefont {J.}~\bibnamefont {Wauters}},\ and\ \bibinfo {author}
  {\bibfnamefont {J.}~\bibnamefont {\ifmmode~\dot{Z}\else \.{Z}\fi{}ylicz}},\
  }\href {https://doi.org/10.1103/PhysRevC.52.R2310} {\bibfield  {journal}
  {\bibinfo  {journal} {Phys. Rev. C}\ }\textbf {\bibinfo {volume} {52}},\
  \bibinfo {pages} {R2310} (\bibinfo {year} {1995})}\BibitemShut {NoStop}%
\bibitem [{\citenamefont {\ifmmode \check{C}\else
  \v{C}\fi{}elikovi\ifmmode~\acute{c}\else \'{c}\fi{}}\ \emph
  {et~al.}(2016)\citenamefont {\ifmmode \check{C}\else
  \v{C}\fi{}elikovi\ifmmode~\acute{c}\else \'{c}\fi{}}, \citenamefont
  {Lewitowicz}, \citenamefont {Gernh\"auser}, \citenamefont {Kr\"ucken},
  \citenamefont {Nishimura}, \citenamefont {Sakurai}, \citenamefont {Ahn},
  \citenamefont {Baba}, \citenamefont {Blank}, \citenamefont {Blazhev},
  \citenamefont {Boutachkov}, \citenamefont {Browne}, \citenamefont
  {de~France}, \citenamefont {Doornenbal}, \citenamefont {Faestermann},
  \citenamefont {Fang}, \citenamefont {Fukuda}, \citenamefont {Giovinazzo},
  \citenamefont {Goel}, \citenamefont {G\'orska}, \citenamefont {Ilieva},
  \citenamefont {Inabe}, \citenamefont {Isobe}, \citenamefont {Jungclaus},
  \citenamefont {Kameda}, \citenamefont {Kim}, \citenamefont {Kwon},
  \citenamefont {Kojouharov}, \citenamefont {Kubo}, \citenamefont {Kurz},
  \citenamefont {Lorusso}, \citenamefont {Lubos}, \citenamefont {Moschner},
  \citenamefont {Murai}, \citenamefont {Nishizuka}, \citenamefont {Park},
  \citenamefont {Patel}, \citenamefont {Rajabali}, \citenamefont {Rice},
  \citenamefont {Schaffner}, \citenamefont {Shimizu}, \citenamefont {Sinclair},
  \citenamefont {S\"oderstr\"om}, \citenamefont {Steiger}, \citenamefont
  {Sumikama}, \citenamefont {Suzuki}, \citenamefont {Takeda}, \citenamefont
  {Wang}, \citenamefont {Watanabe}, \citenamefont {Wu},\ and\ \citenamefont
  {Xu}}]{2016Celikovic}%
  \BibitemOpen
  \bibfield  {author} {\bibinfo {author} {\bibfnamefont {I.}~\bibnamefont
  {\ifmmode \check{C}\else \v{C}\fi{}elikovi\ifmmode~\acute{c}\else
  \'{c}\fi{}}}, \bibinfo {author} {\bibfnamefont {M.}~\bibnamefont
  {Lewitowicz}}, \bibinfo {author} {\bibfnamefont {R.}~\bibnamefont
  {Gernh\"auser}}, \bibinfo {author} {\bibfnamefont {R.}~\bibnamefont
  {Kr\"ucken}}, \bibinfo {author} {\bibfnamefont {S.}~\bibnamefont
  {Nishimura}}, \bibinfo {author} {\bibfnamefont {H.}~\bibnamefont {Sakurai}},
  \bibinfo {author} {\bibfnamefont {D.}~\bibnamefont {Ahn}}, \bibinfo {author}
  {\bibfnamefont {H.}~\bibnamefont {Baba}}, \bibinfo {author} {\bibfnamefont
  {B.}~\bibnamefont {Blank}}, \bibinfo {author} {\bibfnamefont
  {A.}~\bibnamefont {Blazhev}}, \bibinfo {author} {\bibfnamefont
  {P.}~\bibnamefont {Boutachkov}}, \bibinfo {author} {\bibfnamefont
  {F.}~\bibnamefont {Browne}}, \bibinfo {author} {\bibfnamefont
  {G.}~\bibnamefont {de~France}}, \bibinfo {author} {\bibfnamefont
  {P.}~\bibnamefont {Doornenbal}}, \bibinfo {author} {\bibfnamefont
  {T.}~\bibnamefont {Faestermann}}, \bibinfo {author} {\bibfnamefont
  {Y.}~\bibnamefont {Fang}}, \bibinfo {author} {\bibfnamefont {N.}~\bibnamefont
  {Fukuda}}, \bibinfo {author} {\bibfnamefont {J.}~\bibnamefont {Giovinazzo}},
  \bibinfo {author} {\bibfnamefont {N.}~\bibnamefont {Goel}}, \bibinfo {author}
  {\bibfnamefont {M.}~\bibnamefont {G\'orska}}, \bibinfo {author}
  {\bibfnamefont {S.}~\bibnamefont {Ilieva}}, \bibinfo {author} {\bibfnamefont
  {N.}~\bibnamefont {Inabe}}, \bibinfo {author} {\bibfnamefont
  {T.}~\bibnamefont {Isobe}}, \bibinfo {author} {\bibfnamefont
  {A.}~\bibnamefont {Jungclaus}}, \bibinfo {author} {\bibfnamefont
  {D.}~\bibnamefont {Kameda}}, \bibinfo {author} {\bibfnamefont {Y.-K.}\
  \bibnamefont {Kim}}, \bibinfo {author} {\bibfnamefont {Y.~K.}\ \bibnamefont
  {Kwon}}, \bibinfo {author} {\bibfnamefont {I.}~\bibnamefont {Kojouharov}},
  \bibinfo {author} {\bibfnamefont {T.}~\bibnamefont {Kubo}}, \bibinfo {author}
  {\bibfnamefont {N.}~\bibnamefont {Kurz}}, \bibinfo {author} {\bibfnamefont
  {G.}~\bibnamefont {Lorusso}}, \bibinfo {author} {\bibfnamefont
  {D.}~\bibnamefont {Lubos}}, \bibinfo {author} {\bibfnamefont
  {K.}~\bibnamefont {Moschner}}, \bibinfo {author} {\bibfnamefont
  {D.}~\bibnamefont {Murai}}, \bibinfo {author} {\bibfnamefont
  {I.}~\bibnamefont {Nishizuka}}, \bibinfo {author} {\bibfnamefont
  {J.}~\bibnamefont {Park}}, \bibinfo {author} {\bibfnamefont {Z.}~\bibnamefont
  {Patel}}, \bibinfo {author} {\bibfnamefont {M.}~\bibnamefont {Rajabali}},
  \bibinfo {author} {\bibfnamefont {S.}~\bibnamefont {Rice}}, \bibinfo {author}
  {\bibfnamefont {H.}~\bibnamefont {Schaffner}}, \bibinfo {author}
  {\bibfnamefont {Y.}~\bibnamefont {Shimizu}}, \bibinfo {author} {\bibfnamefont
  {L.}~\bibnamefont {Sinclair}}, \bibinfo {author} {\bibfnamefont {P.-A.}\
  \bibnamefont {S\"oderstr\"om}}, \bibinfo {author} {\bibfnamefont
  {K.}~\bibnamefont {Steiger}}, \bibinfo {author} {\bibfnamefont
  {T.}~\bibnamefont {Sumikama}}, \bibinfo {author} {\bibfnamefont
  {H.}~\bibnamefont {Suzuki}}, \bibinfo {author} {\bibfnamefont
  {H.}~\bibnamefont {Takeda}}, \bibinfo {author} {\bibfnamefont
  {Z.}~\bibnamefont {Wang}}, \bibinfo {author} {\bibfnamefont {H.}~\bibnamefont
  {Watanabe}}, \bibinfo {author} {\bibfnamefont {J.}~\bibnamefont {Wu}},\ and\
  \bibinfo {author} {\bibfnamefont {Z.}~\bibnamefont {Xu}},\ }\href
  {https://doi.org/10.1103/PhysRevLett.116.162501} {\bibfield  {journal}
  {\bibinfo  {journal} {Phys. Rev. Lett.}\ }\textbf {\bibinfo {volume} {116}},\
  \bibinfo {pages} {162501} (\bibinfo {year} {2016})}\BibitemShut {NoStop}%
\bibitem [{\citenamefont {Chartier}\ \emph {et~al.}(1996)\citenamefont
  {Chartier}, \citenamefont {Auger}, \citenamefont {Mittig}, \citenamefont
  {L\'epine-Szily}, \citenamefont {Fifield}, \citenamefont {Casandjian},
  \citenamefont {Chabert}, \citenamefont {Ferm\'e}, \citenamefont {Gillibert},
  \citenamefont {Lewitowicz}, \citenamefont {Mac~Cormick}, \citenamefont
  {Moscatello}, \citenamefont {Odland}, \citenamefont {Orr}, \citenamefont
  {Politi}, \citenamefont {Spitaels},\ and\ \citenamefont
  {Villari}}]{1996Chartier}%
  \BibitemOpen
  \bibfield  {author} {\bibinfo {author} {\bibfnamefont {M.}~\bibnamefont
  {Chartier}}, \bibinfo {author} {\bibfnamefont {G.}~\bibnamefont {Auger}},
  \bibinfo {author} {\bibfnamefont {W.}~\bibnamefont {Mittig}}, \bibinfo
  {author} {\bibfnamefont {A.}~\bibnamefont {L\'epine-Szily}}, \bibinfo
  {author} {\bibfnamefont {L.~K.}\ \bibnamefont {Fifield}}, \bibinfo {author}
  {\bibfnamefont {J.~M.}\ \bibnamefont {Casandjian}}, \bibinfo {author}
  {\bibfnamefont {M.}~\bibnamefont {Chabert}}, \bibinfo {author} {\bibfnamefont
  {J.}~\bibnamefont {Ferm\'e}}, \bibinfo {author} {\bibfnamefont
  {A.}~\bibnamefont {Gillibert}}, \bibinfo {author} {\bibfnamefont
  {M.}~\bibnamefont {Lewitowicz}}, \bibinfo {author} {\bibfnamefont
  {M.}~\bibnamefont {Mac~Cormick}}, \bibinfo {author} {\bibfnamefont {M.~H.}\
  \bibnamefont {Moscatello}}, \bibinfo {author} {\bibfnamefont {O.~H.}\
  \bibnamefont {Odland}}, \bibinfo {author} {\bibfnamefont {N.~A.}\
  \bibnamefont {Orr}}, \bibinfo {author} {\bibfnamefont {G.}~\bibnamefont
  {Politi}}, \bibinfo {author} {\bibfnamefont {C.}~\bibnamefont {Spitaels}},\
  and\ \bibinfo {author} {\bibfnamefont {A.~C.~C.}\ \bibnamefont {Villari}},\
  }\href {https://doi.org/10.1103/PhysRevLett.77.2400} {\bibfield  {journal}
  {\bibinfo  {journal} {Phys. Rev. Lett.}\ }\textbf {\bibinfo {volume} {77}},\
  \bibinfo {pages} {2400} (\bibinfo {year} {1996})}\BibitemShut {NoStop}%
\bibitem [{\citenamefont {Commara}\ \emph {et~al.}(2000)\citenamefont
  {Commara}, \citenamefont {del Campo}, \citenamefont {D'Onofrio},
  \citenamefont {Gadea}, \citenamefont {Glogowski}, \citenamefont
  {Jarillo-Herrero}, \citenamefont {Belcari}, \citenamefont {Borcea},
  \citenamefont {{de Angelis}}, \citenamefont {Fahlander}, \citenamefont
  {G\'{o}rska}, \citenamefont {Grawe}, \citenamefont {Hellstr\'{o}m},
  \citenamefont {Kirchner}, \citenamefont {Rejmund}, \citenamefont {Roca},
  \citenamefont {Roeckl}, \citenamefont {Romano}, \citenamefont {Rykaczewski},
  \citenamefont {Schmidt},\ and\ \citenamefont {Terrasi}}]{2000Commara}%
  \BibitemOpen
  \bibfield  {author} {\bibinfo {author} {\bibfnamefont {M.~L.}\ \bibnamefont
  {Commara}}, \bibinfo {author} {\bibfnamefont {J.~G.}\ \bibnamefont {del
  Campo}}, \bibinfo {author} {\bibfnamefont {A.}~\bibnamefont {D'Onofrio}},
  \bibinfo {author} {\bibfnamefont {A.}~\bibnamefont {Gadea}}, \bibinfo
  {author} {\bibfnamefont {M.}~\bibnamefont {Glogowski}}, \bibinfo {author}
  {\bibfnamefont {P.}~\bibnamefont {Jarillo-Herrero}}, \bibinfo {author}
  {\bibfnamefont {N.}~\bibnamefont {Belcari}}, \bibinfo {author} {\bibfnamefont
  {R.}~\bibnamefont {Borcea}}, \bibinfo {author} {\bibfnamefont
  {G.}~\bibnamefont {{de Angelis}}}, \bibinfo {author} {\bibfnamefont
  {C.}~\bibnamefont {Fahlander}}, \bibinfo {author} {\bibfnamefont
  {M.}~\bibnamefont {G\'{o}rska}}, \bibinfo {author} {\bibfnamefont
  {H.}~\bibnamefont {Grawe}}, \bibinfo {author} {\bibfnamefont
  {M.}~\bibnamefont {Hellstr\'{o}m}}, \bibinfo {author} {\bibfnamefont
  {R.}~\bibnamefont {Kirchner}}, \bibinfo {author} {\bibfnamefont
  {M.}~\bibnamefont {Rejmund}}, \bibinfo {author} {\bibfnamefont
  {V.}~\bibnamefont {Roca}}, \bibinfo {author} {\bibfnamefont {E.}~\bibnamefont
  {Roeckl}}, \bibinfo {author} {\bibfnamefont {M.}~\bibnamefont {Romano}},
  \bibinfo {author} {\bibfnamefont {K.}~\bibnamefont {Rykaczewski}}, \bibinfo
  {author} {\bibfnamefont {K.}~\bibnamefont {Schmidt}},\ and\ \bibinfo {author}
  {\bibfnamefont {F.}~\bibnamefont {Terrasi}},\ }\href
  {https://doi.org/https://doi.org/10.1016/S0375-9474(99)00814-3} {\bibfield
  {journal} {\bibinfo  {journal} {Nuclear Physics A}\ }\textbf {\bibinfo
  {volume} {669}},\ \bibinfo {pages} {43} (\bibinfo {year} {2000})}\BibitemShut
  {NoStop}%
\bibitem [{\citenamefont {Elomaa}\ \emph {et~al.}(2009)\citenamefont {Elomaa},
  \citenamefont {Eronen}, \citenamefont {Hager}, \citenamefont {Hakala},
  \citenamefont {Jokinen}, \citenamefont {Kankainen}, \citenamefont {Moore},
  \citenamefont {Rahaman}, \citenamefont {Rissanen}, \citenamefont {Rubchenya},
  \citenamefont {Weber},\ and\ \citenamefont {Aysto}}]{2009Elomaa}%
  \BibitemOpen
  \bibfield  {author} {\bibinfo {author} {\bibfnamefont {V.~V.}\ \bibnamefont
  {Elomaa}}, \bibinfo {author} {\bibfnamefont {T.}~\bibnamefont {Eronen}},
  \bibinfo {author} {\bibfnamefont {U.}~\bibnamefont {Hager}}, \bibinfo
  {author} {\bibfnamefont {J.}~\bibnamefont {Hakala}}, \bibinfo {author}
  {\bibfnamefont {A.}~\bibnamefont {Jokinen}}, \bibinfo {author} {\bibfnamefont
  {A.}~\bibnamefont {Kankainen}}, \bibinfo {author} {\bibfnamefont {I.~D.}\
  \bibnamefont {Moore}}, \bibinfo {author} {\bibfnamefont {S.}~\bibnamefont
  {Rahaman}}, \bibinfo {author} {\bibfnamefont {J.}~\bibnamefont {Rissanen}},
  \bibinfo {author} {\bibfnamefont {V.}~\bibnamefont {Rubchenya}}, \bibinfo
  {author} {\bibfnamefont {C.}~\bibnamefont {Weber}},\ and\ \bibinfo {author}
  {\bibfnamefont {J.}~\bibnamefont {Aysto}},\ }\href
  {https://doi.org/10.1140/epja/i2008-10732-1} {\bibfield  {journal} {\bibinfo
  {journal} {The European Physical Journal A}\ }\textbf {\bibinfo {volume}
  {40}},\ \bibinfo {pages} {1} (\bibinfo {year} {2009})}\BibitemShut {NoStop}%
\bibitem [{\citenamefont {Auranen}\ \emph {et~al.}(2018)\citenamefont
  {Auranen}, \citenamefont {Seweryniak}, \citenamefont {Albers}, \citenamefont
  {Ayangeakaa}, \citenamefont {Bottoni}, \citenamefont {Carpenter},
  \citenamefont {Chiara}, \citenamefont {Copp}, \citenamefont {David},
  \citenamefont {Doherty}, \citenamefont {Harker}, \citenamefont {Hoffman},
  \citenamefont {Janssens}, \citenamefont {Khoo}, \citenamefont {Kuvin},
  \citenamefont {Lauritsen}, \citenamefont {Lotay}, \citenamefont {Rogers},
  \citenamefont {Sethi}, \citenamefont {Scholey}, \citenamefont {Talwar},
  \citenamefont {Walters}, \citenamefont {Woods},\ and\ \citenamefont
  {Zhu}}]{2018Auranen}%
  \BibitemOpen
  \bibfield  {author} {\bibinfo {author} {\bibfnamefont {K.}~\bibnamefont
  {Auranen}}, \bibinfo {author} {\bibfnamefont {D.}~\bibnamefont {Seweryniak}},
  \bibinfo {author} {\bibfnamefont {M.}~\bibnamefont {Albers}}, \bibinfo
  {author} {\bibfnamefont {A.~D.}\ \bibnamefont {Ayangeakaa}}, \bibinfo
  {author} {\bibfnamefont {S.}~\bibnamefont {Bottoni}}, \bibinfo {author}
  {\bibfnamefont {M.~P.}\ \bibnamefont {Carpenter}}, \bibinfo {author}
  {\bibfnamefont {C.~J.}\ \bibnamefont {Chiara}}, \bibinfo {author}
  {\bibfnamefont {P.}~\bibnamefont {Copp}}, \bibinfo {author} {\bibfnamefont
  {H.~M.}\ \bibnamefont {David}}, \bibinfo {author} {\bibfnamefont {D.~T.}\
  \bibnamefont {Doherty}}, \bibinfo {author} {\bibfnamefont {J.}~\bibnamefont
  {Harker}}, \bibinfo {author} {\bibfnamefont {C.~R.}\ \bibnamefont {Hoffman}},
  \bibinfo {author} {\bibfnamefont {R.~V.~F.}\ \bibnamefont {Janssens}},
  \bibinfo {author} {\bibfnamefont {T.~L.}\ \bibnamefont {Khoo}}, \bibinfo
  {author} {\bibfnamefont {S.~A.}\ \bibnamefont {Kuvin}}, \bibinfo {author}
  {\bibfnamefont {T.}~\bibnamefont {Lauritsen}}, \bibinfo {author}
  {\bibfnamefont {G.}~\bibnamefont {Lotay}}, \bibinfo {author} {\bibfnamefont
  {A.~M.}\ \bibnamefont {Rogers}}, \bibinfo {author} {\bibfnamefont
  {J.}~\bibnamefont {Sethi}}, \bibinfo {author} {\bibfnamefont
  {C.}~\bibnamefont {Scholey}}, \bibinfo {author} {\bibfnamefont
  {R.}~\bibnamefont {Talwar}}, \bibinfo {author} {\bibfnamefont {W.~B.}\
  \bibnamefont {Walters}}, \bibinfo {author} {\bibfnamefont {P.~J.}\
  \bibnamefont {Woods}},\ and\ \bibinfo {author} {\bibfnamefont
  {S.}~\bibnamefont {Zhu}},\ }\href
  {https://doi.org/10.1103/PhysRevLett.121.182501} {\bibfield  {journal}
  {\bibinfo  {journal} {Phys. Rev. Lett.}\ }\textbf {\bibinfo {volume} {121}},\
  \bibinfo {pages} {182501} (\bibinfo {year} {2018})}\BibitemShut {NoStop}%
\bibitem [{\citenamefont {Oishi}\ \emph {et~al.}(2018)\citenamefont {Oishi},
  \citenamefont {Fortunato},\ and\ \citenamefont {Vitturi}}]{2018Oishi_LA}%
  \BibitemOpen
  \bibfield  {author} {\bibinfo {author} {\bibfnamefont {T.}~\bibnamefont
  {Oishi}}, \bibinfo {author} {\bibfnamefont {L.}~\bibnamefont {Fortunato}},\
  and\ \bibinfo {author} {\bibfnamefont {A.}~\bibnamefont {Vitturi}},\ }\href
  {https://doi.org/10.1088/1361-6471/aad8f8} {\bibfield  {journal} {\bibinfo
  {journal} {Journal of Physics G: Nuclear and Particle Physics}\ }\textbf
  {\bibinfo {volume} {45}},\ \bibinfo {pages} {105101} (\bibinfo {year}
  {2018})}\BibitemShut {NoStop}%
\bibitem [{\citenamefont {Oishi}(2018)}]{2018Oishi}%
  \BibitemOpen
  \bibfield  {author} {\bibinfo {author} {\bibfnamefont {T.}~\bibnamefont
  {Oishi}},\ }\href {https://doi.org/10.1103/PhysRevC.97.024314} {\bibfield
  {journal} {\bibinfo  {journal} {Phys. Rev. C}\ }\textbf {\bibinfo {volume}
  {97}},\ \bibinfo {pages} {024314} (\bibinfo {year} {2018})}\BibitemShut
  {NoStop}%
\bibitem [{\citenamefont {Chabanat}\ \emph {et~al.}(1997)\citenamefont
  {Chabanat}, \citenamefont {Bonche}, \citenamefont {Haensel}, \citenamefont
  {Meyer},\ and\ \citenamefont {Schaeffer}}]{1997Chabanat_SLY4_v1}%
  \BibitemOpen
  \bibfield  {author} {\bibinfo {author} {\bibfnamefont {E.}~\bibnamefont
  {Chabanat}}, \bibinfo {author} {\bibfnamefont {P.}~\bibnamefont {Bonche}},
  \bibinfo {author} {\bibfnamefont {P.}~\bibnamefont {Haensel}}, \bibinfo
  {author} {\bibfnamefont {J.}~\bibnamefont {Meyer}},\ and\ \bibinfo {author}
  {\bibfnamefont {R.}~\bibnamefont {Schaeffer}},\ }\href
  {https://doi.org/https://doi.org/10.1016/S0375-9474(97)00596-4} {\bibfield
  {journal} {\bibinfo  {journal} {Nuclear Physics A}\ }\textbf {\bibinfo
  {volume} {627}},\ \bibinfo {pages} {710} (\bibinfo {year}
  {1997})}\BibitemShut {NoStop}%
\bibitem [{\citenamefont {Chabanat}\ \emph {et~al.}(1998)\citenamefont
  {Chabanat}, \citenamefont {Bonche}, \citenamefont {Haensel}, \citenamefont
  {Meyer},\ and\ \citenamefont {Schaeffer}}]{1998Chabanat_SLY4_v2}%
  \BibitemOpen
  \bibfield  {author} {\bibinfo {author} {\bibfnamefont {E.}~\bibnamefont
  {Chabanat}}, \bibinfo {author} {\bibfnamefont {P.}~\bibnamefont {Bonche}},
  \bibinfo {author} {\bibfnamefont {P.}~\bibnamefont {Haensel}}, \bibinfo
  {author} {\bibfnamefont {J.}~\bibnamefont {Meyer}},\ and\ \bibinfo {author}
  {\bibfnamefont {R.}~\bibnamefont {Schaeffer}},\ }\href
  {https://doi.org/https://doi.org/10.1016/S0375-9474(98)00180-8} {\bibfield
  {journal} {\bibinfo  {journal} {Nuclear Physics A}\ }\textbf {\bibinfo
  {volume} {635}},\ \bibinfo {pages} {231} (\bibinfo {year} {1998})},\ \bibinfo
  {note} {with erratum: Nuclear Physics A, volume 643, pages 441
  (1998).}\BibitemShut {Stop}%
\bibitem [{\citenamefont {Bartel}\ \emph {et~al.}(1982)\citenamefont {Bartel},
  \citenamefont {Quentin}, \citenamefont {Brack}, \citenamefont {Guet},\ and\
  \citenamefont {Håkansson}}]{SKMS}%
  \BibitemOpen
  \bibfield  {author} {\bibinfo {author} {\bibfnamefont {J.}~\bibnamefont
  {Bartel}}, \bibinfo {author} {\bibfnamefont {P.}~\bibnamefont {Quentin}},
  \bibinfo {author} {\bibfnamefont {M.}~\bibnamefont {Brack}}, \bibinfo
  {author} {\bibfnamefont {C.}~\bibnamefont {Guet}},\ and\ \bibinfo {author}
  {\bibfnamefont {H.-B.}\ \bibnamefont {Håkansson}},\ }\href
  {https://doi.org/http://dx.doi.org/10.1016/0375-9474(82)90403-1} {\bibfield
  {journal} {\bibinfo  {journal} {Nuclear Physics A}\ }\textbf {\bibinfo
  {volume} {386}},\ \bibinfo {pages} {79 } (\bibinfo {year}
  {1982})}\BibitemShut {NoStop}%
\bibitem [{\citenamefont {Vautherin}\ and\ \citenamefont
  {Brink}(1972)}]{72Vautherin}%
  \BibitemOpen
  \bibfield  {author} {\bibinfo {author} {\bibfnamefont {D.}~\bibnamefont
  {Vautherin}}\ and\ \bibinfo {author} {\bibfnamefont {D.~M.}\ \bibnamefont
  {Brink}},\ }\href {https://doi.org/10.1103/PhysRevC.5.626} {\bibfield
  {journal} {\bibinfo  {journal} {Phys. Rev. C}\ }\textbf {\bibinfo {volume}
  {5}},\ \bibinfo {pages} {626} (\bibinfo {year} {1972})}\BibitemShut {NoStop}%
\bibitem [{\citenamefont {Thompson}\ \emph {et~al.}(1977)\citenamefont
  {Thompson}, \citenamefont {Lemere},\ and\ \citenamefont {Tang}}]{77Thom}%
  \BibitemOpen
  \bibfield  {author} {\bibinfo {author} {\bibfnamefont {D.}~\bibnamefont
  {Thompson}}, \bibinfo {author} {\bibfnamefont {M.}~\bibnamefont {Lemere}},\
  and\ \bibinfo {author} {\bibfnamefont {Y.}~\bibnamefont {Tang}},\ }\href
  {https://doi.org/http://dx.doi.org/10.1016/0375-9474(77)90007-0} {\bibfield
  {journal} {\bibinfo  {journal} {Nuclear Physics A}\ }\textbf {\bibinfo
  {volume} {286}},\ \bibinfo {pages} {53 } (\bibinfo {year}
  {1977})}\BibitemShut {NoStop}%
\bibitem [{\citenamefont {Bertsch}\ and\ \citenamefont
  {Esbensen}(1991)}]{1991BE}%
  \BibitemOpen
  \bibfield  {author} {\bibinfo {author} {\bibfnamefont {G.}~\bibnamefont
  {Bertsch}}\ and\ \bibinfo {author} {\bibfnamefont {H.}~\bibnamefont
  {Esbensen}},\ }\href
  {https://doi.org/http://dx.doi.org/10.1016/0003-4916(91)90033-5} {\bibfield
  {journal} {\bibinfo  {journal} {Annals of Physics}\ }\textbf {\bibinfo
  {volume} {209}},\ \bibinfo {pages} {327 } (\bibinfo {year}
  {1991})}\BibitemShut {NoStop}%
\bibitem [{\citenamefont {Esbensen}\ \emph {et~al.}(1997)\citenamefont
  {Esbensen}, \citenamefont {Bertsch},\ and\ \citenamefont
  {Hencken}}]{1997EBH}%
  \BibitemOpen
  \bibfield  {author} {\bibinfo {author} {\bibfnamefont {H.}~\bibnamefont
  {Esbensen}}, \bibinfo {author} {\bibfnamefont {G.~F.}\ \bibnamefont
  {Bertsch}},\ and\ \bibinfo {author} {\bibfnamefont {K.}~\bibnamefont
  {Hencken}},\ }\href {https://doi.org/10.1103/PhysRevC.56.3054} {\bibfield
  {journal} {\bibinfo  {journal} {Phys. Rev. C}\ }\textbf {\bibinfo {volume}
  {56}},\ \bibinfo {pages} {3054} (\bibinfo {year} {1997})}\BibitemShut
  {NoStop}%
\bibitem [{\citenamefont {Hagino}\ and\ \citenamefont {Sagawa}(2005)}]{2005HS}%
  \BibitemOpen
  \bibfield  {author} {\bibinfo {author} {\bibfnamefont {K.}~\bibnamefont
  {Hagino}}\ and\ \bibinfo {author} {\bibfnamefont {H.}~\bibnamefont
  {Sagawa}},\ }\href {https://doi.org/10.1103/PhysRevC.72.044321} {\bibfield
  {journal} {\bibinfo  {journal} {Phys. Rev. C}\ }\textbf {\bibinfo {volume}
  {72}},\ \bibinfo {pages} {044321} (\bibinfo {year} {2005})}\BibitemShut
  {NoStop}%
\bibitem [{\citenamefont {Bertulani}\ and\ \citenamefont
  {S.~Hussein}(2007)}]{07Bertulani_76}%
  \BibitemOpen
  \bibfield  {author} {\bibinfo {author} {\bibfnamefont {C.~A.}\ \bibnamefont
  {Bertulani}}\ and\ \bibinfo {author} {\bibfnamefont {M.}~\bibnamefont
  {S.~Hussein}},\ }\href {https://doi.org/10.1103/PhysRevC.76.051602}
  {\bibfield  {journal} {\bibinfo  {journal} {Phys. Rev. C}\ }\textbf {\bibinfo
  {volume} {76}},\ \bibinfo {pages} {051602} (\bibinfo {year}
  {2007})}\BibitemShut {NoStop}%
\bibitem [{\citenamefont {Tamagaki}(1968)}]{1968Tamagaki}%
  \BibitemOpen
  \bibfield  {author} {\bibinfo {author} {\bibfnamefont {R.}~\bibnamefont
  {Tamagaki}},\ }\href {https://doi.org/10.1143/PTP.39.91} {\bibfield
  {journal} {\bibinfo  {journal} {Progress of Theoretical Physics}\ }\textbf
  {\bibinfo {volume} {39}},\ \bibinfo {pages} {91} (\bibinfo {year}
  {1968})}\BibitemShut {NoStop}%
\bibitem [{\citenamefont {Tamagaki}(1970)}]{1970Tamagaki}%
  \BibitemOpen
  \bibfield  {author} {\bibinfo {author} {\bibfnamefont {R.}~\bibnamefont
  {Tamagaki}},\ }\href {https://doi.org/10.1143/PTP.44.905} {\bibfield
  {journal} {\bibinfo  {journal} {Progress of Theoretical Physics}\ }\textbf
  {\bibinfo {volume} {44}},\ \bibinfo {pages} {905} (\bibinfo {year}
  {1970})}\BibitemShut {NoStop}%
\bibitem [{\citenamefont {Suzuki}(1982)}]{1982Suzuki_NPA}%
  \BibitemOpen
  \bibfield  {author} {\bibinfo {author} {\bibfnamefont {T.}~\bibnamefont
  {Suzuki}},\ }\href
  {https://doi.org/https://doi.org/10.1016/0375-9474(82)90559-0} {\bibfield
  {journal} {\bibinfo  {journal} {Nuclear Physics A}\ }\textbf {\bibinfo
  {volume} {379}},\ \bibinfo {pages} {110} (\bibinfo {year}
  {1982})}\BibitemShut {NoStop}%
\bibitem [{\citenamefont {Suzuki}\ and\ \citenamefont
  {Kohno}(1982)}]{1982Suzuki_PTP}%
  \BibitemOpen
  \bibfield  {author} {\bibinfo {author} {\bibfnamefont {T.}~\bibnamefont
  {Suzuki}}\ and\ \bibinfo {author} {\bibfnamefont {M.}~\bibnamefont {Kohno}},\
  }\href {https://doi.org/10.1143/PTP.68.690} {\bibfield  {journal} {\bibinfo
  {journal} {Progress of Theoretical Physics}\ }\textbf {\bibinfo {volume}
  {68}},\ \bibinfo {pages} {690} (\bibinfo {year} {1982})}\BibitemShut
  {NoStop}%
\bibitem [{\citenamefont {Oishi}\ and\ \citenamefont {Paar}(2019)}]{2019OP_M1}%
  \BibitemOpen
  \bibfield  {author} {\bibinfo {author} {\bibfnamefont {T.}~\bibnamefont
  {Oishi}}\ and\ \bibinfo {author} {\bibfnamefont {N.}~\bibnamefont {Paar}},\
  }\href {https://doi.org/10.1103/PhysRevC.100.024308} {\bibfield  {journal}
  {\bibinfo  {journal} {Phys. Rev. C}\ }\textbf {\bibinfo {volume} {100}},\
  \bibinfo {pages} {024308} (\bibinfo {year} {2019})}\BibitemShut {NoStop}%
\bibitem [{\citenamefont {Misra}\ and\ \citenamefont
  {Sudarshan}(1977)}]{1977Misra}%
  \BibitemOpen
  \bibfield  {author} {\bibinfo {author} {\bibfnamefont {B.}~\bibnamefont
  {Misra}}\ and\ \bibinfo {author} {\bibfnamefont {E.~C.~G.}\ \bibnamefont
  {Sudarshan}},\ }\href {https://doi.org/https://doi.org/10.1063/1.523304}
  {\bibfield  {journal} {\bibinfo  {journal} {Journal of Mathematical Physics}\
  }\textbf {\bibinfo {volume} {18}},\ \bibinfo {pages} {756} (\bibinfo {year}
  {1977})}\BibitemShut {NoStop}%
\bibitem [{\citenamefont {Chiu}\ \emph {et~al.}(1977)\citenamefont {Chiu},
  \citenamefont {Sudarshan},\ and\ \citenamefont {Misra}}]{1977Chiu}%
  \BibitemOpen
  \bibfield  {author} {\bibinfo {author} {\bibfnamefont {C.~B.}\ \bibnamefont
  {Chiu}}, \bibinfo {author} {\bibfnamefont {E.~C.~G.}\ \bibnamefont
  {Sudarshan}},\ and\ \bibinfo {author} {\bibfnamefont {B.}~\bibnamefont
  {Misra}},\ }\href {https://doi.org/10.1103/PhysRevD.16.520} {\bibfield
  {journal} {\bibinfo  {journal} {Phys. Rev. D}\ }\textbf {\bibinfo {volume}
  {16}},\ \bibinfo {pages} {520} (\bibinfo {year} {1977})}\BibitemShut
  {NoStop}%
\bibitem [{\citenamefont {Levitan}(1988)}]{1988Levitan}%
  \BibitemOpen
  \bibfield  {author} {\bibinfo {author} {\bibfnamefont {J.}~\bibnamefont
  {Levitan}},\ }\href
  {https://doi.org/https://doi.org/10.1016/0375-9601(88)90329-5} {\bibfield
  {journal} {\bibinfo  {journal} {Physics Letters A}\ }\textbf {\bibinfo
  {volume} {129}},\ \bibinfo {pages} {267} (\bibinfo {year}
  {1988})}\BibitemShut {NoStop}%
\bibitem [{\citenamefont {Ram\'{\i}rez~Jim\'enez}\ and\ \citenamefont
  {Kelkar}(2021)}]{2021Jimenez}%
  \BibitemOpen
  \bibfield  {author} {\bibinfo {author} {\bibfnamefont {D.~F.}\ \bibnamefont
  {Ram\'{\i}rez~Jim\'enez}}\ and\ \bibinfo {author} {\bibfnamefont {N.~G.}\
  \bibnamefont {Kelkar}},\ }\href {https://doi.org/10.1103/PhysRevA.104.022214}
  {\bibfield  {journal} {\bibinfo  {journal} {Phys. Rev. A}\ }\textbf {\bibinfo
  {volume} {104}},\ \bibinfo {pages} {022214} (\bibinfo {year}
  {2021})}\BibitemShut {NoStop}%
\end{thebibliography}
